\documentclass[english]{article}
\usepackage{lmodern}

\usepackage[T1]{fontenc}
\usepackage[latin9]{inputenc}
\usepackage{color}
\usepackage{babel}
\usepackage{verbatim}
\usepackage{varioref}
\usepackage{amsmath}
\usepackage{graphicx}
\usepackage[unicode=true,
 bookmarks=false,
 breaklinks=true,pdfborder={0 0 0},backref=false,colorlinks=false]
 {hyperref}
\hypersetup{pdftitle={Subtyping in Java with Generics and Wildcards is a Fractal},
 pdfauthor={Moez A. AbdelGawad},
 pdfsubject={Subtyping in Generic Nominally-Typed OOP is a Fractal},
 pdfkeywords={Subtyping; Generics; Fractals; Nominal-Typing; Wildcards; Java; C\#; C++; Scala}}

\makeatletter
\numberwithin{equation}{section}
\numberwithin{figure}{section}
\newenvironment{lyxcode}
{\par\begin{list}{}{
\setlength{\rightmargin}{\leftmargin}
\setlength{\listparindent}{0pt}
\raggedright
\setlength{\itemsep}{0pt}
\setlength{\parsep}{0pt}
\normalfont\ttfamily}%
 \item[]}
{\end{list}}
\newcommand{\code}[1]{\texttt{#1}}

\@ifundefined{showcaptionsetup}{}{%
 \PassOptionsToPackage{caption=false}{subfig}}
\usepackage{subfig}
\makeatother

\usepackage{listings}

\begin{document}

\title{}

\title{Subtyping in Java with Generics and Wildcards is a Fractal}

\author{Moez A. AbdelGawad\\
\smallskip{}
College of Mathematics, Hunan University\\
Changsha 410082, Hunan, P.R. China\\
\smallskip{}
Informatics Research Institute, SRTA-City\\
New Borg ElArab, Alexandria, Egypt\\
\smallskip{}
moez@cs.rice.edu}
\maketitle
\begin{abstract}
While developing their software, professional object-oriented (OO)
software developers keep in their minds an image of the subtyping
relation between types in their software. The goal of this paper is
to present an observation about the graph of the subtyping relation
in Java, namely the observation that, after the addition of generics---and
of wildcards, in particular---to Java, the graph of the subtyping
relation is no longer a simple directed-acyclic graph (DAG), as in
pre-generics Java, but is rather a \emph{fractal}. Further, this observation
equally applies to other mainstream nominally-typed OO languages (such
as C\#, C++ and Scala) where generics and wildcards (or some other
form of `variance annotations') are standard features. Accordingly,
the shape of the subtyping relation in these OO languages is more
complex than a tree or a simple DAG, and indeed is also a fractal.
Given the popularity of fractals, the fractal observation may help
OO software developers keep a useful and intuitive mental image of
their software's subtyping relation, even if it is a little more frightening,
and more amazing one than before. With proper support from IDEs, the
fractal observation can help OO developers in resolving type errors
they may find in their code in lesser time, and with more confidence.
\end{abstract}

\section{Introduction}

For helping themselves in writing, debugging and maintaining their
software, professional software developers using object-oriented programming
languages keep in their minds an image or picture of the subtyping
relation between types in their software while they are developing
their software. In pre-generics Java~\cite{JLS05}, the number of
possible object types (also called \emph{reference types}) for a fixed
set of classes in a program was a \emph{finite} number (however large
it was), and, more importantly, the structure of the subtyping relation
between these types (and hence of the mental image a developer kept
in mind) was simple: the graph of the subtyping relation between classes
and interfaces (\emph{i.e.}, with multiple-inheritance of interfaces)
was a simple directed-acyclic graph (DAG), and the graph of the subtyping
relation between classes alone (\emph{i.e.}, with single-inheritance
only, more accurately called the \emph{subclassing} relation) was
simply a tree.\footnote{Trees are well-known data structures, and a DAG, in short, is a generalization
of a tree where a node is further allowed to have more than one parent
node (\emph{i.e.}, not just one parent as in a tree) but the node
cannot be a parent of itself, even if indirectly; a DAG can thus have
no cycles, hence being `acyclic'.} This fact about the graph of the subtyping relation applies not only
to Java but, more generally, also to the non-generic sublanguage of
other mainstream nominally-typed OO languages similar to Java, such
as C\#~\cite{CSharp2007}, C++~\cite{CPP2011}, and Scala~\cite{Odersky09}.

Today, generics and wildcards (or some other form of `variance annotations')
are a standard feature of mainstream nominally-typed OO languages.
The inheritance relation, between classes (and interfaces and traits,
in OO languages that support these notions) is still a finite relation,
and its shape is still the same as before: a simple DAG. But, given
the possibility of arbitrary nesting of generic types, the number
of possible object types in a generic Java program has become infinite,
and the shape of the subtyping relation in nominally-typed OO languages
has become more complex than a tree or a simple DAG.

It is thus natural to wonder, ``\emph{What is the shape of the subtyping
relation in Java}, now after the addition of generics and wildcards?''

This question on subtyping in Java is similar to one Benoit Mandelbrot,
in the 1960s, wondered about: ``How long is the coast of Britain?''.
At that time, some mathematicians (including many computer scientists)
used to believe that mathematics was perfect because it had completely
banished pictures, even from elementary textbooks. Mandelbrot, using
computers, put the pictures back in mathematics, by discovering fractals,
and, in the process, finding that Britain's coast has infinite length.

The goal of this paper is to present and defend, even if incompletely
and unconventionally (using mainly hierarchy diagrams, and only using
equations suggestively), a fundamental observation about the graph
of the subtyping relation in Java. We observed that, after the addition
of generics---and of wildcards, in particular---to Java, the graph
of the subtyping relation is still a DAG, but is no longer a simple
DAG but is rather one whose structure can be better understood, of
all possibilities, also as a \emph{fractal} -- and in fact, as we
explain below, an intricately constructed fractal (albeit a different
kind of fractal than that of Britain's coast).

To motivate our observation, we use very simple generic class declarations
to present in the paper some diagrams for the subtyping relation that
represent the iterative construction of the subtyping graph, in the
hope of making the construction process very simple to understand
and thus make the fractal observation very clear. To further argue
for and strengthen the observation, we also suggest algebraic equations
for mathematically describing the subtyping fractal and its construction
process. (Our equations are akin of recursive domain equations of domain theory
that are used to construct `reflexive domains'. The similarity is
suggestive of a strong relationship, possibly even suggesting reflexive
domains---useful for giving mathematical meaning for programming languages---might
be fractals too, even though we refrain from arguing for this claim
here.)

\subsection{Practical Significance}

Given the popularity fractals enjoy nowadays, we believe the fractal
observation about subtyping in nominally-typed OO languages may help
OO software developers keep a useful and intuitive mental image of
their software's subtyping relation, even if it is a little more
frightening, and more amazing one than the one they had before. As
an immediate application of the fractal observation, IDEs (Integrated
Development Environments) that OO developers use can make developers'
lives easier, making them develop their software faster and with more
confidence, by presenting to them parts of the fractal representing
the subtyping relation in their software and allowing developers to
``zoom-in''/``zoom-out'' on sections of the fractal/relation that
are of interest to the developers, in order for them to better understand
the typing relations in their software and so that they may resolve
any type errors in their code more quickly and more confidently. OO
language designers may also benefit from the fractal observation,
since having a better understanding of the subtyping relation may
enable them to have a better understanding of the interactions between
different features of OO languages---such as the three-tiered interaction,
in Java, between generics (including wildcard types), `lambdas'
(formerly known as `closures') and type inference---leading designers
to improve the design of the language, and to better design and implement
its compilers. Finally, in allusion to Joshua Bloch's well-known
quote when considering adding closures to Java, we hope, by making
the fractal observation about subtyping, to enable decreasing (or
at least, more accurately estimating) the ``\emph{complexity budget}''
paid for adding generics and wildcards to Java.

\section{The Fractal Observation}

As any standard definition (or an image) of a fractal will reveal,
fractals (sometimes also called \emph{recursive graphs}, or \emph{self-referential
graphs}) are drawings or graphs that are characterized by having ``minicopies''
of themselves inside of them~\cite{mandelbrot1977fractals,barnsley2013fractals,Fractals2011,WikipediaFractal2014}.
Given their \emph{self-similar} nature, when zooming in on a fractal
it is not a surprise to find a copy of the original fractal spring
up. More generally, the minicopy is not an exact copy, but some \emph{transformation}
of the original: it may be the original rotated, translated, reflected,
and so on. As such, when constructing a fractal iteratively (as is
standard) it is also not a surprise to add details to the construction
of the fractal by using (transformed) copies of the fractal as constructed
so far (\emph{i.e.}, as it exists in the current iteration of the
construction) to get a better, more accurate approximation of the
final fractal (See Figure~\ref{fig:Fractals}, and~\cite{WikipediaFractal2014}).

While it may not be immediately obvious to the unsuspecting, but ``having
transformed minicopies of itself'' is exactly what we have noticed
also happens in (the graph of) the subtyping relation of Java---and
of other similar generic nominally-typed OO languages such as C\#,
C++, and Scala---after generics and wildcards were added to the Java
type system.%
{} Figure~\vref{fig:First-Iterations} presents a drawing of the first
steps in the construction of a subtyping graph, to illustrate and
give a ``flavor'' of the observation.%
{} In Section~\ref{sec:Observation-Illustrated}, to motivate presenting
the subsequent \emph{transformations observation} in Section~\ref{sec:Transformations-Observation},
we present a more precise and more detailed diagram---one that, unlike
Figure~\ref{fig:First-Iterations}, uses no `raw types',%
{} and has an additional class \code{D}.

\noindent \begin{center}
\noindent \begin{center}
\begin{figure}[h]
\noindent \begin{centering}
\includegraphics{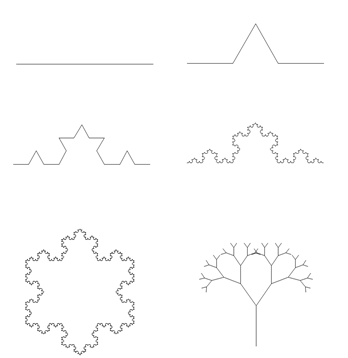}
\par\end{centering}

\protect\caption{\label{fig:Fractals}Fractals: (First Steps in Constructing) The Koch
Curve and (a Step in Constructing) a Fractal Tree}
\end{figure}

\par\end{center}
\par\end{center}

\noindent \begin{center}
\begin{figure*}[h]
\noindent \begin{centering}
\includegraphics[scale=0.44]{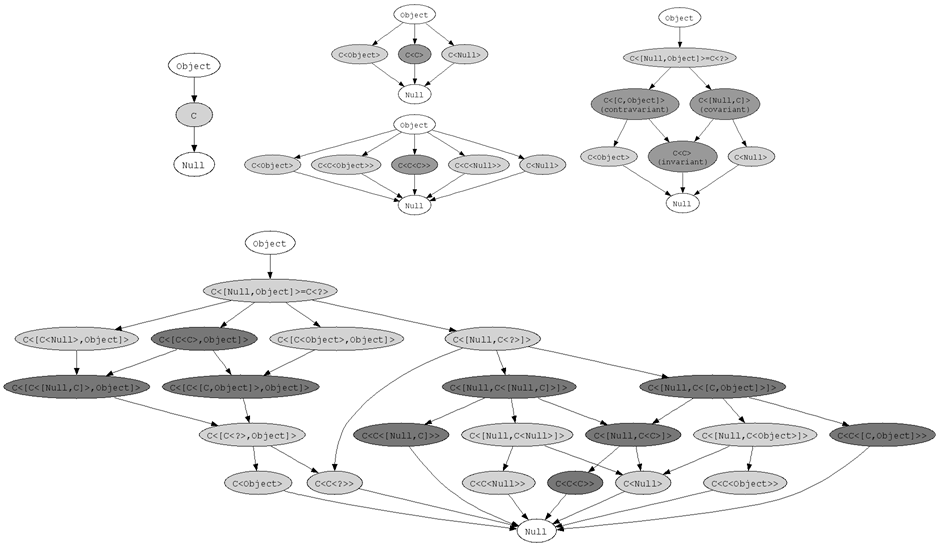}
\par\end{centering}

\protect\caption{\label{fig:First-Iterations}First Iterations of Constructing a Subtyping
Graph in Java}
\end{figure*}

\par\end{center}

\section{\label{sec:Observation-Illustrated}Observation Illustrated: Con\-struct\-ing
The Subtyping Fractal}

\subsection{Subclassing}

To illustrate our main observation and how the subtyping fractal is
constructed, let us assume we have the non-generic class \code{Object}
(which extends/subclasses no other classes, \emph{i.e.}, is at the
top of the subclassing/inheritance hierarchy), and that we have, as
expressed in the two simple lines of code below, two generic classes
\code{C} and \code{D} that extend class \code{Object} and that
take one (unbounded) type parameter. Similarly, and crucial to seeing
the subtyping graph as a fractal, we also assume we have a ``hidden''
(\emph{i.e.}, inexpressible in some OO languages, such as Java) non-generic
class \code{Null} at the bottom of the class inheritance hierarchy
(whose only instance is the \code{null} object, which in Java is
an instance of every class and can be assigned to a variable of any
object type\footnote{Even when the \code{\textbf{instanceof}} operator in Java, only for
developers' convenience, returns \code{false} as the value of the
expression `\code{null \textbf{instanceof} C}', for any class \code{C}.
 Incidentally, the possibility of having \emph{non-nullable} classes,
existing in some OO languages such as C\#, may need some special provision
to envision the resulting subtyping graph, such as providing an additional
class \code{Empty} that extends class \code{Null} and has no instances.
For domain-theorists: The `class type' corresponding then to class
\code{Empty} will be the empty object domain, \emph{i.e.}, the domain
whose only ``instance'', or member, is $\bot_{O}$, ``the non-terminating
object''.)}.
\begin{lyxcode}
\begin{lstlisting}
class C<T> extends Object {}
class D<T> extends Object {}
\end{lstlisting}

\end{lyxcode}
Figure~\vref{fig:Fig1} demonstrates the subclassing hierarchy (\emph{a.k.a.},
inheritance hierarchy) based on assuming these class declarations.

\noindent \begin{center}
\begin{figure}[h]
\noindent \centering{}\subfloat[\label{fig:Fig1}Subclassing]{\protect\begin{centering}
\protect\includegraphics[scale=0.5]{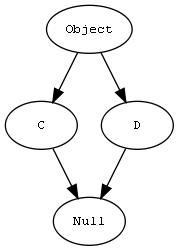}\protect
\par\end{centering}

}~~~~~~~~~\subfloat[\label{fig:Fig2}Subtyping (Nest. Level 0)]{\protect\centering{}\protect\includegraphics[scale=0.37]{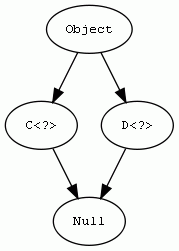}\protect}
\end{figure}

\par\end{center}

\subsection{From Subclassing to Subtyping}

\subsubsection{Nesting Level 0}

The declared inheritance relation between class (and interface/trait)
names in a program is the \emph{starting point}\footnote{Or `the initial graph' or `the base step' of the recursion ``crank,''
or the `skeleton'~\cite{GEB} of the graph, or the `seed'~\cite{bird1998introduction}
of the fractal.} for constructing the graph of the subtyping relation in nominally-typed
OO languages, including Java (note the use of the identification of
type inheritance and subtyping in nominally-typed OOP~\cite{NOOP,NOOPbook,NOOPsumm,OOPOverview13,InhSubtyNWPT13,AbdelGawad2015,AbdelGawad2015a}
to interpret `class extension' as `subtyping between corresponding
class types'. We discuss the role of nominality in more detail in
Section~\ref{sec:NomVsStruct}). Figure~\vref{fig:Fig2} shows that
the ``default type argument'', namely \code{?} (the unbounded wildcard
type), is used in this initial step as the type argument for all generic
classes to form type names for corresponding class types.

\subsubsection{Nesting Level 1 ... and Beyond}

\begin{figure*}[p]
\noindent \begin{centering}
\includegraphics[scale=0.33]{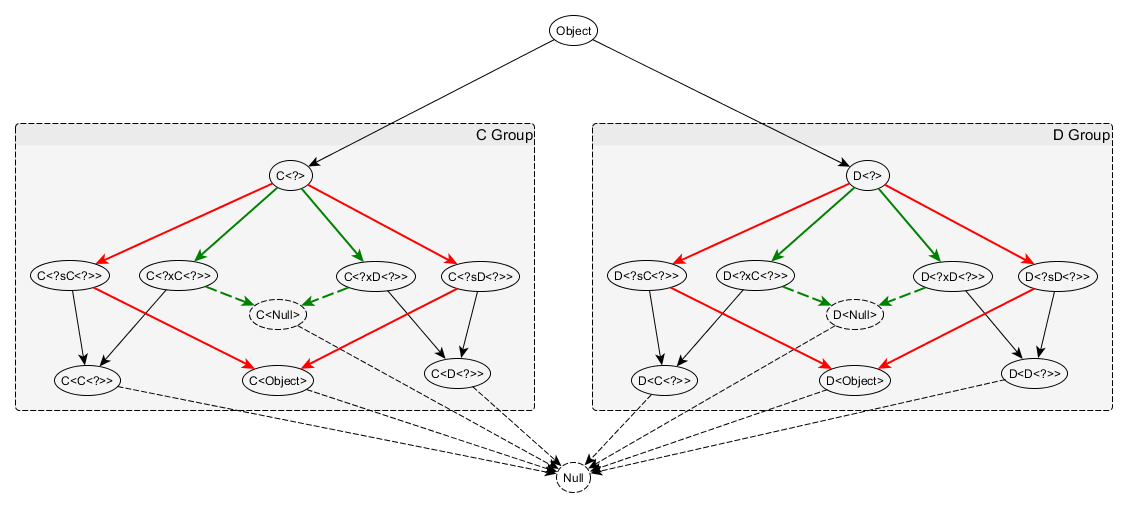}
\par\end{centering}

\protect\caption{\label{fig:Fig3}Subtyping: (First) Inductive Step. Nesting Level
1}
\end{figure*}

Figure~\vref{fig:Fig3} demonstrates how the (names of) types in
the next iteration of constructing the subtyping fractal (\emph{i.e.},
of the iteration numbered $i+1$, which we can ``see'' after looking
at iteration $i$ if we ``zoom in'' one step) constructing are constructed
by replacing/substituting all the \code{?}'s in level/iteration
0 (the base step) with \emph{three} different forms of each type \code{T}
in the previous level (level $i$), namely \code{?~extends~T} (covariance),
\code{?~super~T} (contravariance), and \code{T} (invariance).%
\footnote{Which corresponds to defining the fractal using the equation $G=G_{0}(I(G))$
(see below). Replacing each of the \emph{innermost} (or, all?) \code{?}s
of a type (``holes'' in the type) in level $i$ with a \code{\#}
(a hash, as a placeholder), then replacing these \code{\#}s with
three different forms of each one of the types in the previous level
(level $i$), or in level 0, to construct names of the types of the
new level (corresponding to equation $G=G(I(G))$ or $G=G(I(G_{0}))$).
See further comments below for a note on the likely equality of the
first two equations, and on the likely uselessness of the third equation
as defining the fractal.}

Covariant, contravariant and invariant subtyping rules are then used
to decide the subtyping relation between all the newly constructed
types (note that, due to the inclusion of types \code{Object} and
\code{Null} in level 0 and in all subsequent levels, all level $i$
types are \emph{also} types of level/iteration $i+1$. This motivates
the notion of the rank of a type. The level/iteration in which a type
\emph{first} appears is called the \emph{rank} of the type. As such,
types \code{Object} and \code{Null} are always of rank 0). In Figure~\ref{fig:Fig3}
we use \code{?xT} and \code{?sT} as short-hands for \code{? extends T}
and \code{? super T} respectively.

\paragraph{The effect of variant subtyping rules on the subtyping graph:}
\begin{itemize}
\item Covariant Subtyping: The level 0 graph is \emph{copied} inside \code{C<?>}
and \code{D<?>} (illustrated by \textbf{\textcolor{green}{green}}
arrows in diagrams). For ten types \code{T} (from 2 non-generic classes
+ 2 generic classes $\times$ 4 types in level 0), we have paths\\
\code{Object -> C<?~extends~T> -> Null}, and\\
 \code{Object -> D<?~extends~T> -> Null}\\
(Note: \code{?~extends~Null} is the same as \code{Null}. Inexpressible
in Java).
\item Contravariant Subtyping: The level 0 graph is \emph{flipped} (turned
upside-down) inside \code{C<?>} and \code{D<?>} (illustrated by
\textbf{\textcolor{red}{red}} arrows in diagrams). For ten types \code{T},
like for covariance, we have paths\\
\code{Object -> C<?~super~T> -> Null}, and\\
\code{Object -> D<?~super~T> -> Null}\\
(Note: \code{?~super~Object} is the same as \code{Object}. See footnote~\vref{fn:(A-bug-in}
regarding current Java behavior).
\item Invariant Subtyping: The level 0 graph is \emph{flattened} inside
\code{C<?>} and \code{D<?>} (no corresponding arrows in diagrams).
For ten types \code{T}, like for covariance, we have paths\\
\code{Object -> C<T> -> Null}, and\\
\code{Object -> D<T> -> Null} .
\end{itemize}

\subsection{Type Intervals}

\begin{figure*}[p]
\noindent \begin{centering}
\includegraphics[scale=0.33]{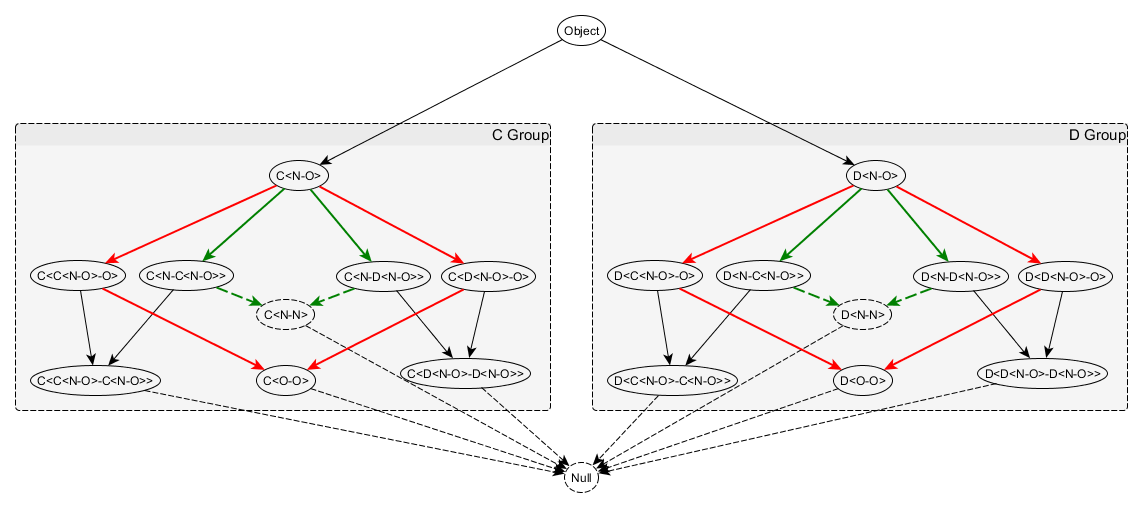}
\par\end{centering}

\protect\caption{\label{fig:Fig4}Subtyping Generalization with Intervals. Nesting
Level 1}
\end{figure*}

Figure~\vref{fig:Fig4} illustrates how to use the notion of type
intervals~\cite{AbdelGawad2016a,AbdelGawad2016c} to combine all
three (\emph{i.e.}, covariant, contravariant and invariant) subtyping
rules (and to add even more types to the subtyping relation in later
iterations/nesting levels).

In Figure~\ref{fig:Fig4}, we have all three transformations applied
to level 0 graph and embedded inside \code{C<?>} and \code{D<?>}
(note that bounds of an interval can degenerately be equal types,
corresponding to invariance). For twenty types \code{S} and \code{T}
(where \code{S} is a subtype of \code{T} in the previous iteration/level),
from 2 non-generic classes + 2 generic classes $\times$ 9 intervals
in level 0, we have \code{Object -> C<S-T> -> Null}, \code{Object -> D<S-T> -> Null}
(The notation \code{S-T} means the interval with lowerbound \code{S}
and upperbound \code{T}. For brevity, we use \code{O} for \code{Object}
and \code{N} for \code{Null} ). If class \code{C} or class \code{D}
had subclasses other than \code{Null}, this graph diagram would have
been even richer---\emph{i.e.}, it would have had more types---than
the graph in Figure~\vref{fig:Fig3}. (It can be noted that the \code{Null}
type is useful in expressing intervals. Yet the diagram can be presented
without it, using \code{extends} only or \code{super} only, while
allowing but not requiring a naked \code{?}; or, for brevity, using
a symbol like \code{<:}).

\medskip{}

Note: The types \code{Null}, \code{C<Null>}, and \code{D<Null>},
inside dotted graph nodes in Figure~\vref{fig:Fig3} and Figure~\vref{fig:Fig4},
are currently \emph{inexpressible} in Java (\emph{i.e.}, as of Java
8, based on the assumption that these types are of little practical
use to developers.) Subtyping relations involving these inexpressible
types are also currently of little use to Java developers (except
in type inference). Accordingly, they also are drawn in Figure~\ref{fig:Fig4}
using dotted graph edges.\footnote{\label{fn:(A-bug-in}(\textbf{A bug in Javac}): Also, as of Java 8,
we have noted that Java does not currently identify \code{?~super~Object}
with \code{Object}, and as such a variable \code{b} of type \code{C<?~super~Object>},
for example, \emph{cannot} be assigned to a variable \code{a} of
type \code{C<Object>} (\emph{i.e.}, for the statement \code{a=b;}
the Java compiler \code{javac} currently emits a type error with
an unhelpful semi-cryptic error message that involves `wildcard capturing')
even as Java allows the opposite assignment of \code{a} to \code{b}
(\emph{i.e.}, the statement \code{b=a;}), implying that, even though
Java currently correctly sees \code{C<Object>} as a subtype of \code{C<?~super~Object>},
it currently does \emph{not} consider \code{C<?~super~Object>} as
a subtype of \code{C<Object>}. Given that there are no supertypes
of type \code{Object} (the class type corresponding to class \code{Object}),
and it is not expected there will ever be any, we believe the Java
type system should be fixed to identify the two type arguments \code{?~super~Object}
and \code{Object}, and thus correctly allow the mentioned currently-disallowed
assignments.}

\medskip{}

It should now be clear how to constructing the rest of the subtyping
fractal. Each next nesting level of generics corresponds to ``zooming
one level in'' in the subtyping fractal, and the construction of
the new ``zoomed-in'' graph is done using the same method above,
where wildcards (or, intervals) over the previous subtyping graph
substitute all the \code{?} in that graph to produce the next level
graph of the subtyping relation. And there is nothing in generics
that disallows arbitrarily-deep, potentially infinite, nesting.

\section{\label{sec:Transformations-Observation}The Transformations Observation}

While making the fractal observation, we made yet another observation
that helps explain the fractal observation more deeply. In particular,
we noted that in constructing the graph of the subtyping relation,
when moving from types of a specific level of nesting to types of
the next deeper level (\emph{i.e.}, when ``zooming in'' inside the
graph of the relation, or when doing the inductive step of the recursive
definition of the graph), \emph{three} kinds of \emph{transformations}
are applied to the level $i$ subtyping graph, in agreement with the
general nature of fractals having transformed minicopies of themselves
embedded within. We call these three transformations the \emph{identity}
(or, \emph{copying}) transformation, the \emph{upside-down }reflection
(or, \emph{flipping }transformation), and the \emph{flattening} transformation.
The first transformation (identity) makes an exact copy of the input
subtyping relation, the second transformation (upside-down reflection)
flips over the relation (a subtype in the input relation becomes a
supertype, and vice versa), while the third transformation ``attempts
to do both (\emph{i.e.}, the identity and flipover transformations),''
in effect making types that were related in its input subtyping relation
be \emph{unrelated} in its output subtyping relation (hence the output
of this transformation is a ``flat'' relation, called an \emph{anti-chain}.)

Explaining this observation regarding the subtyping fractal in terms
of OO subtyping is done by noting that the three mentioned transformations
correspond to (in fact, result from) the covariant subtyping rule,
contravariant subtyping rule, and invariant subtyping rule, respectively.
This is demonstrated, in a very abridged manner, in Figure~\vref{fig:Fig3}
(with the green arrows corresponding to copying the previous level
graph, corresponding to covariant subtyping, the red arrows corresponding
to flipping over the previous level graph, corresponding to contravariant
subtyping.) It should be noted that \emph{also} the level 1 graph
as a \emph{whole} is the same structure as the level 0 graph when
the `C Group' nodes are lumped into one node and the same for the
`D Group' node. That means that, in agreement with the graph being
a fractal (where self-similarity must exist at all levels of \emph{scale}),
when the graph of subtyping is ``viewed from far'' it \emph{looks
the same} as the level 0 graph. In fact, when looked at from a far
enough distance this similarity to the level 0 graph will be the case
for all level $i$ , where $i\geq1$, graphs.

\section{\label{sec:NomVsStruct}Nominally-typed OOP vs. Struct\-urally-typed
OOP}

It should be noted that class names information (\emph{a.k.a.}, nominality,
and `nominal type information') of nominally-typed OO languages
(such as Java, C\#, C++, and Scala) is used in the base/first construction
step in constructing the subtyping relation between generic types
as a fractal. In contrast, structurally-typed OO languages (such as
OCaml~\cite{OCamlWebsite}, Moby~\cite{Fisher1999}, PolyTOIL~\cite{Bruce2003},
and StrongTalk~\cite{Bracha1993}), known mainly among programming
languages researchers, do \emph{not }have such a simple base step,
since a record type corresponding to a class (with at least one method)
in these languages does \emph{not} have a finite number of supertypes
to begin with, given that ``superclasses of a class'' in the program,
when viewed structurally as supertypes of record types, do \emph{not}
form a finite set. Any record type has an infinite set of record subtypes
(due to their width-subtyping rule~\cite{TAPL}). Accordingly, a
record type with a method---\emph{i.e.}, a member having a function
type---causes the record type to have an infinite set of \emph{supertypes},
due to contravariance of the type of the method. Adding-in a depth-subtyping
rule makes the subtyping relation between record types with functional
member types even more complex.

This motivates suspecting that subtyping in structurally-typed OO
language is a \emph{dense} relation, in which every pair of non-equal
types in the relation has a third type, not equal to either member
of the pair, that is ``in the middle'' between the two elements
of the pair, \emph{i.e.}, that is a subtype of the supertype (the
upperbound) of the pair and a supertype of the subtype (the lowerbound)
of the pair. In fact this may turn out to be simple to prove. Due
to a class in generic nominally-typed OO languages having a finite
set of superclasses in the subclassing relation, subtyping in generic
nominally-typed OO languages languages is not an (everywhere) dense
relation, and the subclassing relation in these languages forms a
simple finite basis (the ``skeleton'') for constructing the subtyping
relation. For structurally-typed OO languages (where record types
with functional members are a must, to model structural objects),
this basis (the ``skeleton'') is infinite and thus the ``fractal''
structural subtyping graph (if indeed it is a fractal) is not easy
to draw or to even imagine.

For more details on the technical and mathematical differences between
nominally-typed and structurally-typed OOP, the interested reader
may consult~\cite{Findler04,Malayeri08,Ostermann08,NOOP,NOOPbook,OOPOverview13,NOOPsumm,InhSubtyNWPT13,AbdelGawad2015a,AbdelGawad2015,AbdelGawad2016,AbdelGawad2016a}.

\section{Conclusion}

In this paper we presented an observation connecting subtyping in
generic nominally-typed OO languages to fractals. We presented diagram
for graphs of the subtyping relation demonstrating the iterative process
of constructing the relation as a fractal. We further made an observation
connecting the three variant subtyping rules in generic OOP to the
three transformations done on the graph of the relation for embedding
inside the relation. We further noted some possible differences between
generic nominally-typed OOP and polymorphic structurally-typed OOP
as to the fractal nature of their subtyping relations. (See the Appendix
for some further notes, observations and conclusions that may be built
on top of the observations and discussions we made, including a suggestive
discussion on the use of algebraic equations to precisely describe
the generic OO subtyping relation as a fractal).

\bibliographystyle{plain}

\appendix

\section{Further Notes}

The following notes and observations can be added to the ones we made
in the main paper:
\begin{enumerate}
\item \noindent Relations on Type Intervals: For type intervals $I=[S,T]$,
where $S<:T$, with lowerbound ($S$) and upperbound ($T$), two relations
on intervals can be defined that can help in constructing the subtyping
fractal: An interval containing another interval (the \emph{contains}
relation: $S_{1}<:S_{2}\wedge T_{2}<:T_{1}$), and an interval preceding
another interval (the \emph{precedes} relation: $T_{1}<:S_{2}$).
\item \noindent The \emph{Pruning Transformation}: Bounds, \emph{i.e.},
lowerbounds or upperbounds, on a type parameter limit (\emph{i.e.},
decrease) the types of level $i$ that can substitute the holes (the
\code{?}s) when constructing a type in level $i+1$, so pruning means
that a substitution \emph{respects} these declared bounds.
\item \noindent Demonstration Software: An interactive Mathematica program
that demonstrates the iterative construction of the subtyping hierarchy,
for multiple simple class hierarchies, up to four nesting levels is
available upon request (The program uses the \code{Manipulate} function
of Mathematica 6, is formatted as a Mathematica 6 demo, and is in
the Mathematica .nb format, \emph{i.e.}, the file format Mathematica
has used as of 2007.)
\item \noindent Multi-arity: Generic classes with \emph{multiple} type parameters
simply result in types with multiple ``holes'' at the same nesting
level for the same class.
\item \noindent Graph Matrices: Representing successive subtyping graphs
as adjacency matrices (0-1 matrices) is useful in computing (paths
in graph of) the relation (and in computing containment of intervals).
(Using $(I-A)^{-1}$, with binary addition and multiplication of matrices,
to compute the transitive closure of the relation and thus paths/intervals
over it).
\item \noindent Category Theory: Given the use of the notion of \emph{operads}
in category theory to model self-similarity~\cite{spivak2014category},
we intend to consider the possibility of using\emph{ }operads to express
and communicate the fractal nature of the generic OO subtyping relation.
\item \noindent Algebraic Equations: According to Benoit Mandelbrot, Hermann
Well wrote that `the angel of geometry and the devil of algebra share
the stage, illustrating the difficulties of both.' Turning to some
algebra, we expect the graph of the subtyping relation to be described
by a recursive equation, as is the case for many fractals. We anticipate
this equation to be (something along the lines of) 
\[
G=G_{0}(copy(G)+flip(G)+flatten(G)),
\]
where $G_{0}$ stands for the initial graph (the `skeleton' of the
subtyping fractal, resulting from turning the subclassing relation
into a subtyping relation by using \code{?} as the default type argument
for generic classes), and the application of $G_{0}$ to its argument
(another graph) means the \emph{substitution} (similar to $\beta$\nobreakdash-reduction
in $\lambda$\nobreakdash-calculus) of its ``holes'' (the \code{?}'s
in its types/nodes) with the argument graph (\emph{i.e.}, the graph
$copy(G)+flip(G)+flatten(G)$ which applies the three above-mentioned
transformations to $G$, and where $+$ means ``subtyping-respecting
union'' of component graphs.) 
\end{enumerate}
\noindent \begin{enumerate}
\item \noindent More on Algebraic Equations:

\begin{enumerate}
\item \noindent The $G_{0}$ in the equation\emph{ }above\emph{---i.e.},
the graph of the first iteration of the subtying relation, which is
directly based on the subclassing relation---is what makes (all iterations/approximations
of) the graph $G$ have the same structure ``when viewed from far'',
\emph{i.e.}, when zooming out of it, as the subclassing relation).
\item \noindent To construct approximations of $G$ iteratively, the equation
can be interpreted to mean
\[
G_{i+1}=G_{0}(copy(G_{i})+flip(G_{i})+flatten(G_{i})),
\]
which means when constructing approximations to $G$ we construct
elements of the sequence 
\[
G_{0}=G_{0},
\]
\[
G_{1}=G_{0}(copy(G_{0})+flip(G_{0})+flatten(G_{0})),
\]
\begin{eqnarray*}
G_{2} & = & G_{0}(copy(G_{1})+flip(G_{1})+flatten(G_{1}))\\
 & = & G_{0}(copy(G_{0}(copy(G_{0})+flip(G_{0})+flatten(G_{0})))+\\
 &  & flip(G_{0}(copy(G_{0})+flip(G_{0})+flatten(G_{0})))+\\
 &  & flatten(G_{0}(copy(G_{0})+flip(G_{0})+flatten(G_{0}))))
\end{eqnarray*}
\[
G_{3}=G_{0}(copy(G_{2})+flip(G_{2})+flatten(G_{2}))=...,
\]
... etc.
\item \noindent Another seemingly-equivalent recursive equation for describing
the subtyping graph $G$ is
\[
G=G(copy(G_{0})+flip(G_{0})+flatten(G_{0})),
\]
which, even though not in the more familiar $x=f(x)$ format, has
the advantage of showing that $G$ (the limit, infinite graph) is
equivalent to (isomorphic to) substituting its own holes with transformations
of $G_{0}$, \emph{i.e.}, that the substitution does \emph{not} affect
the final infinite graph $G$ (just as adding 1 to $\omega$, the
limit of natural numbers, does not affect its cardinality; $|\omega|=|\omega+1|$.)
It also reflects the zooming-in fact (opposite to the zooming-out
fact above) that when zooming-in into $G$ we find (transformed copies
of) $G_{0}$ each time we zoom in, ad infinitum. (See Note~\ref{enu:More-Levels/Iterations:-To}
below for why we believe this third equation may in fact be \emph{incorrect}.)\\

\end{enumerate}
\item \noindent Algebraic Equations with Intervals: With intervals, the
equation above becomes simpler and more general, where, if $I$ is
the function computing all the intervals over a graph, we then have
\[
G=G_{0}(I(G))\textrm{, or, }G=G(I(G_{0})),
\]
or, most accurately,
\[
G=G(I(G)).
\]
Note that the three equations agree on defining $G_{1}=G_{0}(I(G_{0}))$.
The three equations disagree however on later terms of the construction
sequence. They, for example, define $G_{2}=G_{0}(I(G_{1}))$, $G_{2}=G_{1}(I(G_{0}))$,
and $G_{2}=G_{1}(I(G_{1}))$, respectively. The equivalence of the
three equations (\emph{i.e.}, of the resulting graph from each) is
unlikely, but a mathematical proof or a convincing intuitive proof
of that is needed (see Note~\ref{enu:More-Levels/Iterations:-To}
below, however).\\

\item \noindent Benefits and Applications: An obvious benefit of the observation
in this paper is to demonstrate one more (unexpected?) place where
fractals show up. Yet an additional benefit, and practical application,
of the observation may be to apply some of the theory developed for
fractals to better the understanding of the subtyping relation in
OO languages, possibly leading to providing a better understanding
of their generic type systems and thus developing better OO language
compilers.
\item \noindent Parameterizing classes \code{Object} and \code{Null}:
At least one needs to be non-parameterized, if not both? Otherwise
we may have an unbounded infinite ascending chain of supertypes (see
Section~\ref{sec:NomVsStruct}.) (What will then be the meaning of
\code{?}, and be the default type argument?)%

\item \noindent \label{enu:More-Levels/Iterations:-To}More Levels/Iterations:
To further demonstrate the fractal observation, and to help resolve
which of the three equations above (best) describes the graph of the
subtyping relation, we draw the level 2 graph $G_{2}$ using a simpler
initial graph (\emph{i.e.}, the `skeleton') than we used for the
earlier figures. See Figure~\vref{fig:SubtypingC01} and Figure~\vref{fig:SubtypingC12b}.\\
\begin{figure}
\noindent \centering{}\includegraphics[bb=0bp 0bp 378bp 361bp,scale=0.5]{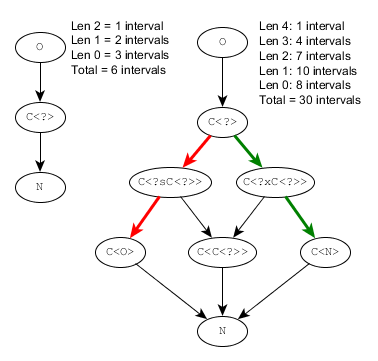}\protect\caption{\label{fig:SubtypingC01}Subtyping levels 0, 1 with one generic class
(\protect\code{C}) }
\end{figure}
\begin{figure*}
\includegraphics[scale=0.5]{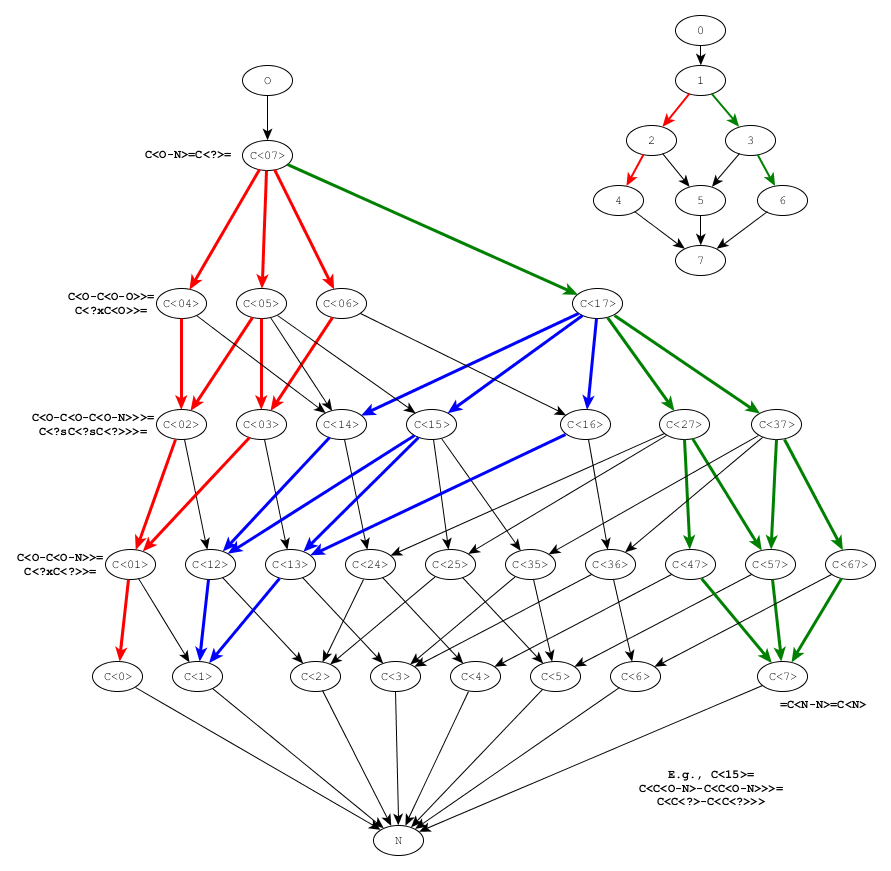}\protect\caption{\label{fig:SubtypingC12b}Subtyping levels 1, 2, corresponding to
equation $G=G_{0}(I(G))$ and Bounded Contravariance/Graph-Flip (Highlighted
in Blue)}
\end{figure*}
\\
Some notes on $G_{0}$, $G_{1}$, $G_{2}$:

\begin{enumerate}
\item $G_{2}$ , constructed as $G_{0}(I(G_{1}))$, has 32 nodes, and 66
edges.
\item The number of levels in graphs $G_{0}$, $G_{1}$, $G_{2}$, ... (\emph{i.e.},
the maximum path length) increases by two each time (2, 4, 6, 8, ...).
This is clear in the diagrams, particularly ones with colored arrows.
\item The number of nodes and edges in $G_{0}$, $G_{1}$, $G_{2}$, ...:

\begin{enumerate}
\item Nodes: $G_{0}$\textbf{=3}=(2+1), $G_{1}$\textbf{=8}=(2+3+2+1), $G_{2}$\textbf{=32}=(2+8+10+7+4+1),
... \textbf{???}=(2+32+66+<...4 numbers...>+1).
\item Edges: $G_{0}$\textbf{=2},\textbf{ }$G_{1}$\textbf{=10},\textbf{
}$G_{2}$\textbf{=66}, ... \textbf{???}.
\end{enumerate}
\item Algebraic Equations:

\begin{enumerate}
\item $G_{2}$, as constructed above, is the same graph as $G_{1}(I(G_{1}))$
... !!
\item Thus, $G_{0}(I(G_{1}))=G_{1}(I(G_{1}))$, meaning that, given $G_{1}=G_{0}(I(G_{0}))$,
we have 
\[
G_{0}(I(G_{0}(I(G_{0}))))=G_{0}(I(G_{0}))\left(I(G_{0}(I(G_{0})))\right)
\]

\item The skeptic reader may trying constructing the graph corresponding
to the equation $G_{2}=G_{1}(I(G_{1}))$
\item \textbf{Proof}: Each type/node constructed in $G_{1}(I(G_{1}))$ is
constructed in $G_{0}(I(G_{1}))$ (and vice versa, which is easy to
see).

\begin{enumerate}
\item Same for proving $G_{1}(I(G_{0}))=G_{0}(I(G_{0}))$, which means we
have 
\[
G_{0}(I(G_{0}))\left(I(G_{0})\right)=G_{0}(I(G_{0}))
\]

\end{enumerate}
\item (An Analogy) Something unknown becoming known. Knowing it again does
not add new.
\item Philosophical observation, using `Old' = $G_{i}$, `New' = $G_{i+1}$:

\begin{enumerate}
\item New in New = New in Old
\item Old in New != New in Old
\item Old in Old = New = Old in New
\end{enumerate}
\end{enumerate}
\item In addition to subgraphs highlighted in green and red (which show
an exact copy and a flipped copy, due to covariance and contravariance
respectively) of $G_{1}$ inside $G_{2}$, Figure~\vref{fig:SubtypingC12b}
also shows a miniature \emph{pruned} flipped copy of $G_{1}$ inside
$G_{2}$, highlighted in blue (due to bounded contravariance).\end{enumerate}
\end{enumerate}

\end{document}